\documentclass[10pt,letterpaper]{article}
\usepackage{opex3}
\usepackage{color}
\usepackage{amsmath}
\usepackage{epstopdf}
\usepackage{cite}
\usepackage{subcaption}
\begin{document}

\title{Fault-tolerant and finite-error localization for point emitters within the diffraction limit}

\author{Zong Sheng Tang$^1$, Kadir Durak$^{1,*}$ and Alexander Ling$^{1,2}$}

\address{$^1$Centre for Quantum Technologies, National University of Singapore, Block S15, 3 Science Drive 2, 117543 Singapore
\\
$^2$Department of Physics, National University of Singapore, Block S12, 2 Science Drive 3, 117551 Singapore}




\begin{abstract}
We implement an estimator for determining the separation between two incoherent point sources. This estimator relies on image inversion interferometry and when used with the appropriate data analytics, it yields an estimate of the separation with finite-error, even when the sources come arbitrarily close together. The experimental results show that the technique has a good tolerance to noise and misalignment, making it an interesting consideration for high resolution instruments.
\end{abstract}


\bibliography{library.bib}



\section{Introduction}

The challenge of localization for point emitters within the standard diffraction limit is an ongoing research topic. The limit for resolving the separation of point sources in direct imaging is commonly linked to Rayleigh's criterion which states that two sources of equal intensity are not resolvable when their separation is smaller than the the radius of the first minima of an airy disk~\cite{rayleigh1879}.
This is a simple criterion based on the resolving power of the eye.
An alternative, Sparrow's limit, is based on the separation where the sum of the two Airy profiles no longer provides a saddle~\cite{sparrow1916}, and is the limit for contrast enhancement techniques~\cite{pearlman10}.
These techniques share the common feature that they rely on image-plane photon counting, and suffer from a divergent mean squared error as the separation tends to zero~\cite{bettens99}.
Image processing to determine the mid-point of radiating sources exhibit the same behaviour in the error, although they are able to estimate the mid-point with arbitrary precision.

Modern microscopy techniques avoid the pitfalls of diverging mean squared error, by restricting the emission behaviour of the sources~\cite{hell94,klar00,lanni95,gustafsson99,Biteen2008,Rittweger2009,schroder2010super}. It is interesting, however, to consider how the error can be reduced when the emitters are inaccessible and the emission behaviour cannot be controlled. 
Motivated by quantum estimation theory, it was recently shown to be possible to avoid the problem of diverging error for arbitrarily small separations~\cite{tsang1_15,tsang2_15,nair16}.
In this paper, we report the implementation of the estimator proposed in~\cite{nair16}, which utilizes an image inversion interferometer~\cite{wicker07, wicker09}.
These previous works utilized photon counting.
We experimentally demonstrate that the technique also works with conventional photodetectors, using power in lieu of accumulated photon number.
We show that this estimator, coupled with an appropriate data processing technique ~\cite{nair16}, produces an estimate of the distance between two sources with finite error for arbitrary separation.

\section{The estimator}
\subsection{Image inversion interferometry}
\begin{figure}[htbp]
\centering\includegraphics[width=10cm]{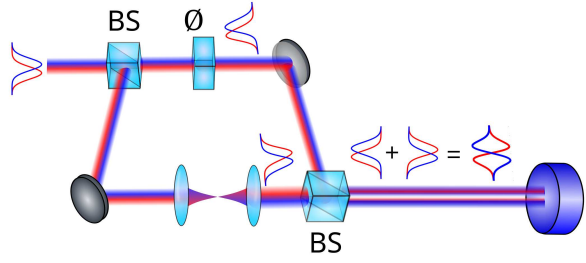}
\caption{The image inversion interferometer is essentially a balanced Mach-Zehnder interferometer with an appropriate delay in one arm, and an image inverter in the other arm. The output power in the arm associated with destructive interference for on-axis light is monitored. This output arm is sensitive to the separation of the point sources. When the sources completely overlap on the optical axis, the power at the image plane is at its weakest value.}
\label{fig:mzi}
\end{figure}

The underlying mechanism of image inversion interferometry is the interference of a beam with its inverted image. The scheme for such an interferometer is provided in Figure~\ref{fig:mzi}. The device is essentially a Mach-Zehnder interferometer with an image inverter in one arm. In our experiment the image inversion is performed using a pair of dispersive lenses, which is sufficient for a proof-of-principle demonstration.

The centroid of two sources is aligned with the central axis of the image inverter. When beams in one path (path $B$) pass through the image inverter, the spatial coordinate of the beams undergo reflection with respect to the centroid. Meanwhile, an optical delay, $\phi$, is inserted in path $A$ to maximize interference at the second beamsplitter. We then monitor the arm that outputs the anti-symmetric component of the combined electric field. This arm is associated with destructive interference for on-axis light, and should be at its minimal power value when the point sources completely overlap on-axis. The output of the interferometer is sensitive to the separation of the point sources.

Following Figure~\ref{fig:mzi}, the electric field for two point emitters are labelled red and blue. Mathematically, the electric field of either emitter in the output plane can be written as
\begin{equation}
E_i(r)=E_{i,A}(r+d)-E_{i,B}(r-d), \quad i\in\{red,blue\},
\end{equation}
where $d$ is the separation of the source from the centroid. The power in the output plane is \begin{equation} P=\int dr |E_{blue}(r)|^2+\int dr |E_{red}(r)|^2 \end{equation} as both red and blue do not interfere with each other and the total power is the sum of individual power.

In our experiment the point sources are obtained from the output of an illuminated single-mode optical fibre. The electric field of the resulting point spread function can be approximated by a Gaussian distribution
\begin{equation}
E(\mathbf{r},\mathbf{r}_0,w)=E_{0}\exp{\left(-\frac{|\mathbf{r}-\mathbf{r}_0|}{2w}\right)^2},
\end{equation}
where $\mathbf{r}$ is the radial position, $\mathbf{r}_0$ is the center of the beam, $E_0$ is the field amplitude and $w$ is the spread of the beams. $w$ can be related to full width at half maximum (FHWM) by FHWM$=4\sqrt{\log 2}w$. When the beams are  interfering destructively with their inverted images, the total power on the image plane is
\begin{equation}
P_{total}=P_A+P_B-2\sqrt{P_A P_B}\exp{\left(-\frac{d^2}{2w^2}\right)}.
\end{equation}
Dividing both sides by ($P_{A}+P_{B}$), we obtain a quantity $\beta$, that is identified as the ratio of residual power after interference,

\begin{align}
\label{betaideal}
\beta=\frac{P_{total}}{P_{A}+P_{B}}=&1-2\frac{\sqrt{P_{ A} P_{B}}}
{P_{A}+P_{B}}\exp{\Bigg(-\frac{d^2}{2w^2}\Bigg)}\\
\approx& 1-\exp{\Bigg(-\frac{d^2}{2w^2}\Bigg)}\Bigg|_{P_{A}\sim P_{B}}.
\end{align}

With $\beta$ as a measured quantity, and $w$ typically known from the optical instrument's performance, it is straightforward to obtain an estimate of the separation, $d_{est}$

\begin{equation}
d_{est}=  w\sqrt{2\log \left(\frac{1}{1-\beta }\right)}~\text{if }~0\leq\beta<1 .
\label{dest}
\end{equation}

\subsection{Error Analysis}
An image inversion interferometer alone is insufficient to prevent the uncertainty in the value of $d_{est}$ from diverging at small separation. Handling the data with conventional error propagation techniques~\cite{ku1966} still leads to divergent behaviour. The general form of the error of a function $f$ assuming independent and identical distribution of variables is
\begin{equation}
(\sigma_f(x_i...x_n))^2=\sum_i^n\left(\frac{\partial f}{\partial x_i}\right)^2(\sigma_{ x_i})^2 .
\label{errorprop}
\end{equation}
Substituting $d_{est}$ into equation~(\ref{errorprop}) gives

\begin{equation}
\sigma^2_{d_{est}}=\bigg|\frac{\sigma_\beta ^2 w ^2}{2 (1-\beta )^2 \log \left(\frac{1}{1-\beta }\right)}\bigg|+\bigg|2 \sigma_w ^2 \log
   \left(\frac{1}{1-\beta }\right)\bigg|.
   \label{divergeerror}
\end{equation}
In equation~(\ref{divergeerror}), the first term diverges when $\beta\rightarrow 0$ or $\beta\rightarrow 1$, which is unsatisfactory.

An alternative technique for estimating uncertainty is to determine directly the root mean square error (RMSE) of the actual distribution in the data. In this approach, the RMSE for a cluster of observed $\beta$ values is computed against a selected mean value.
It can be shown that using this method, the uncertainty in the value of $d_{est}$ does not diverge when separation is small~\cite{nair16}. 
For example, if the RMSE error for $\beta$ is denoted as $\sigma_\beta$, the error in $d_{est}$ is

\begin{equation}
\sigma_{d_{est}}=\sqrt{\int^\infty_{-\infty}d\beta f(\beta)(d_{est}(\beta)-d)^2}.
\end{equation}

\begin{figure}[t]\centering\includegraphics[width=7cm]{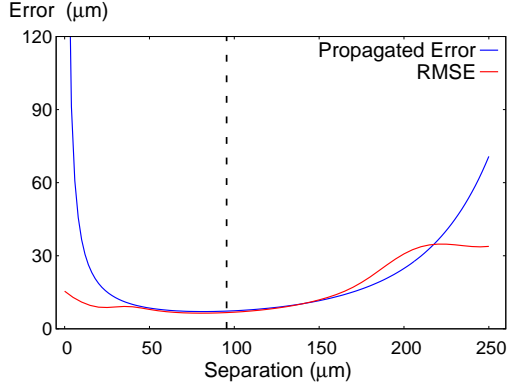}

\caption{Trends in error of estimated distance using conventional error propagation and the root mean square error (RMSE) technique. The vertical line represents the Sparrow limit for our experimental system.
}
\label{errorcomparetheo}
\end{figure}

Error estimations from both methods are plotted in Figure~\ref{errorcomparetheo}, assuming that $\beta$ has a normal distribution $f(\beta)$ whose standard deviation $\sigma_\beta$ value is 0.05. The RMSE technique clearly yields a finite uncertainty even at zero separation. Both techniques have the same performance about the Sparrow limit.

\section{Experiment}

\begin{figure}[t]
\centering
\includegraphics[width=13cm]{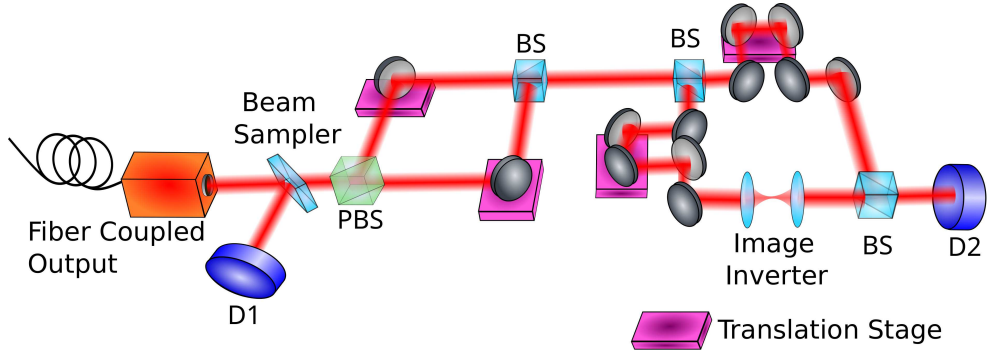}
\caption{ Schematic diagram of the experimental setup. The two test sources, with orthogonal polarizations, are obtained from an illuminated single-mode optical fibre whose output is separated by a polarizing beam splitter (PBS). The separation of the sources is controlled by adjusting mirrors mounted on translation stages.}
\label{setupm}
\end{figure}

The schematic for the experiment is shown in Figure~\ref{setupm}. To prepare two incoherent point sources we used light from a fiber-coupled continuous-wave Helium-Neon laser. The output of the illuminated fiber is split by a polarizing beamsplitter (PBS), and then recombined on a standard (50:50) beamsplitter. The orthogonal polarization between the two light fields prevents mutual interference. By translating the mirrors before the recombining beamsplitter, the separation between the two ``sources" can be arbitrarily adjusted, allowing the interferometer to be calibrated. It should be noted that the technique works even for non-orthogonal polarization, as long as the sources are mutually incoherent.

The image inverter consists of two aspheric lenses in a confocal configuration. Optical trombones allow control of the optical path difference between the arms of the interferometer. The coherence length is experimentally verified to be below 2 mm. The phase of the two paths is adjusted to interfere destructively at the position of the detector D2. 
The value for $w$ in our experiment is determined to be $95.5$~$\pm1.9$~$\mu$m for both arms. The power from each arm reaching detector $D_2$, denoted as $P_A$ and $P_B$, are estimated from the following equation

\begin{equation}
P_{i}=\frac{ <P_{D_{2,i}}> }{<P_{D_1}>} P_{D_1}, \quad i\in\{A,B\}.
\end{equation}
The value of $P_{D_{2,A}}$ is obtained by observing the power reading at $D_2$ when the arm $B$ is blocked, and vice versa.
The error in $P_{i}$ is found to follow a normal distribution with a standard deviation of approximately 1\% about the mean value.

\begin{figure}[htbp]
\centering
\includegraphics[width=11cm]{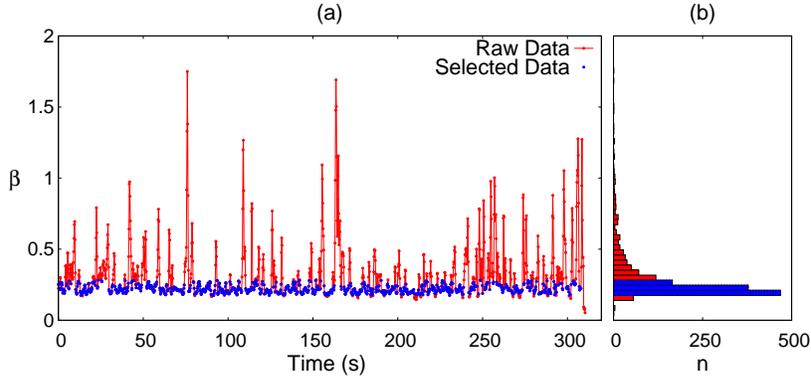}
\caption{(a) Experimental observation of $\beta$ values. Clearly, there is a floor to the value of $\beta$ with interferometer instability causing outliers in the observational data. To filter out the outliers, a quartile sorting technique was adopted with accepted data points presented in blue (color online). (b) The sorted data.}
\label{filter}
\end{figure}

For convenience, the path length difference between the two arms was not locked. Instead, to filter out any outliers in the data caused by fluctuating path length difference, the values for $\beta$ were sorted into quartiles \cite{huberbook,moore2007}. The data corresponding to quartiles $Q_1,~Q_2,~Q_3$ are identified and used to reject data points that do not fulfil the condition
\begin{equation}
|\beta_i-Q_2|<\frac{1}{2}(Q_3-Q_1).
\label{criteria}
\end{equation}
An example of the distribution of raw and selected data is shown in Figure~\ref{filter}. 

\begin{figure}[t]
\centering
\includegraphics[width=14cm]{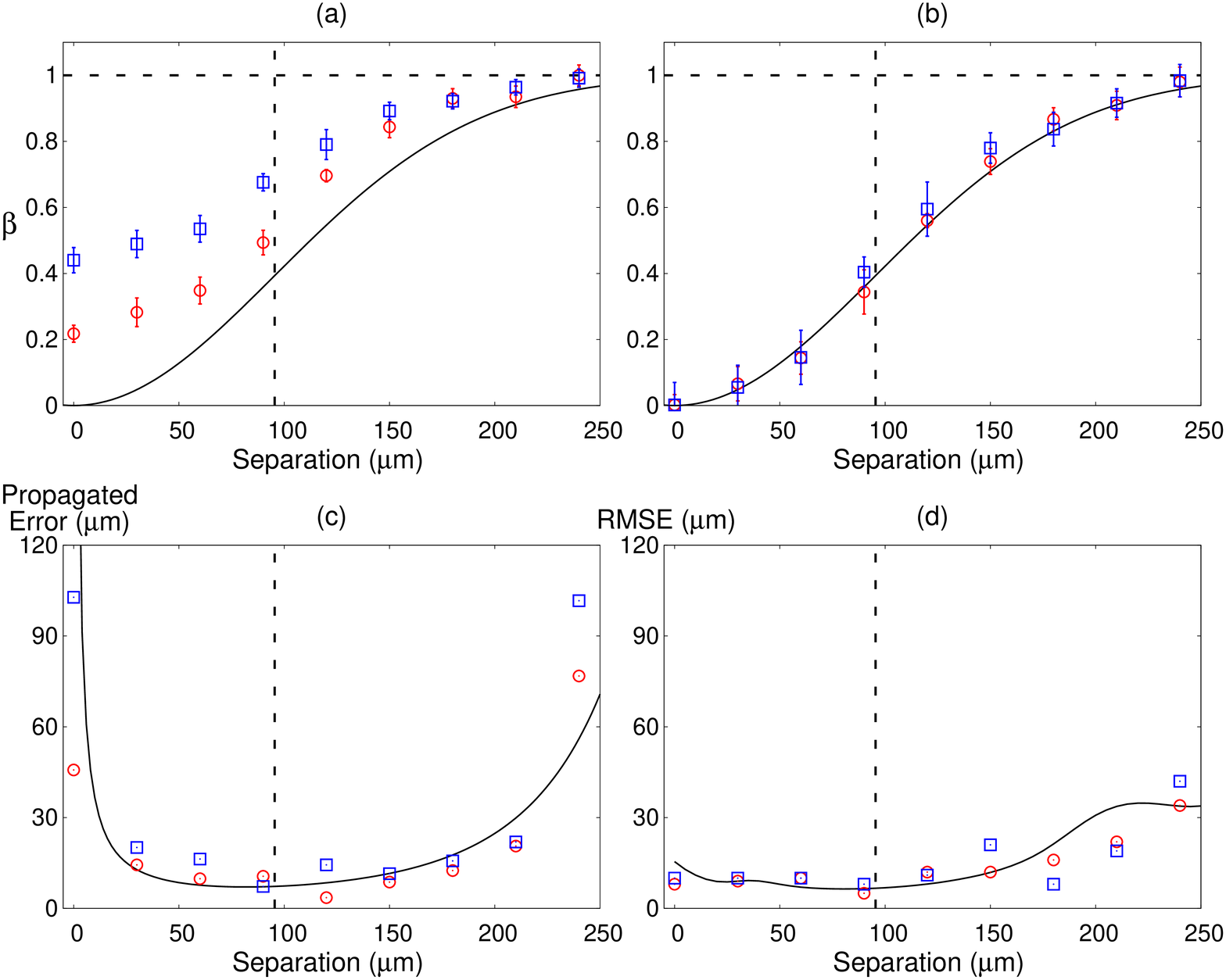}
\caption{ (a) Measured values of residual power $\beta$ against set separation for two different degrees of interference. The blue (red) data points have a minimal value of 0.44 (0.22) when the two point sources overlap. For comparison, the theoretical expectation of $\beta$ for a perfect instrument is provided (solid line). The black vertical line indicates Sparrow's limit in our experiment. (b) The same data points after subtracting for background. These plots serve as a calibration curve for the instrument. (c) The error in estimated separation, when using conventional error propagation, diverges for small separation. (d) The error in estimated separation, when using the RMSE technique, remains finite even for very small separation.}
\label{betavd}
\end{figure}

In principle, the $\beta$ value should reach zero for complete overlap, but instrumentation error prevents the ideal value from being reached. 
Factors contributing to instrumentation error include misalignment between the center of the image inverter and the centroid, and the dispersive properties of the lenses. 
Both result in a minimum finite value for $\beta$ when the two point sources overlap.
This provides a calibration curve to the instrument enabling a unique value of $d_{est}$ to be associated with every value of $\beta$.
This is shown in Figure~\ref{betavd}~(a) for two different floor values for $\beta$.
A further correction can be made, by considering the $\beta$ value when $d$=0~$\mu$m as a constant noise floor to be subtracted
\begin{equation}
\beta=\frac{{P}_{{D}_2}-{F}}
{(P_A+P_B)-{F}},
\label{norm}
\end{equation}
where $F$ is the floor due to instrument error. Implementing this correction results in the data points that follow the theoretical value as shown in Figure~\ref{betavd}~(b).

From the data points observed in Figure~\ref{betavd}, it is possible to derive the associated error in any estimate of separation $d$.
These errors derived using conventional error propagation and the RMSE technique are respectively shown in panels (c) and (d) of Figure~\ref{betavd}, and follow the predicted trends.

\section{Conclusion}
The implementation of a finite-error estimator for the separation of two point sources has been presented. This is achieved by combining the sensitivity of an image inversion interferometer with the RMSE method for estimating error. This localization is accompanied by finite-error in the estimated separation even for very small separation values, and performed without modifying the emission of the sources. In this experiment, the light sources are derived from a single laser.
It is not expected that a change to thermal light sources would fundamentally affect the performance of the instrument. Additional care would be needed to ensure that the short coherence length associated with thermal light can still achieve sufficient interference.

This technique is also robust against instrument error. Noise introduced by random fluctuation in the path lengths can be rejected using the quartile technique. The presence of the floor in the residual power value $\beta$ can be treated as a background that can be subtracted to obtain good agreement with the theoretical behaviour. This fault-tolerant technique for localization of point sources could be of interest when building resolving instruments that operate outside laboratory conditions.

During the preparation of this manuscript, the authors became aware of a general proposal for performing finite-error localization ~\cite{tsang16}. The use of data analytics inspired by a study of quantum estimation theory, as applied in this paper and emerging within the literature, suggests a new capability for optical metrology.\\

\textbf{Acknowledgements} This research is supported by the National Research Foundation, Prime Minister's Office, Singapore (Grant Number: NRF-CRP12-2013-02). James A. Grieve assisted with the manuscript.
\end{document}